\documentclass[11pt]{article}

\usepackage[]{acl}

\usepackage{multirow}
\usepackage{tabularx}

\usepackage{listings}
\usepackage{caption}

\usepackage{times}
\usepackage{latexsym}

\usepackage[T1]{fontenc}

\usepackage[utf8]{inputenc}

\usepackage{microtype}

\usepackage{inconsolata}

\usepackage{graphicx}

\usepackage{booktabs}
\usepackage{amsmath}
\usepackage{xspace}
\usepackage[most]{tcolorbox}
\usepackage{cuted}
\usepackage{framed}
\usepackage{capt-of}

\title{Beyond Correctness: Enhancing Architectural Reasoning in Code LLMs via Scalable Labeling with Agentic Judgment}

\author{
  \parbox{\textwidth}{
  \centering
  Kirill Vasilevski\textsuperscript{1}\thanks{\; Equal contribution},
  Ximing Dong\textsuperscript{1}\footnotemark[1],
  Benjamin Rombaut\textsuperscript{1}\footnotemark[1],
  Milad Soltany\textsuperscript{1}\footnotemark[1],
  Ruochen Deng\textsuperscript{1}\footnotemark[1],
  Jiahuei (Justina) Lin\textsuperscript{1},
  Arthur Leung\textsuperscript{1},
  Dayi Lin\textsuperscript{3},
  Boyuan Chen\textsuperscript{1},
  Shaowei Wang\textsuperscript{2}\thanks{\;\;Corresponding author.},\\
  Ahmed E. Hassan\textsuperscript{3}\\ 
  \textsuperscript{1}Centre for Software Excellence, Huawei Canada \\
  \textsuperscript{2}Department of Computer Science, University of Manitoba, Canada \\
  \textsuperscript{3}School of Computing, Queen's University, Canada \\[-0.1em]
    \scalebox{0.75}{\texttt{\{kirill.vasilevski,ximing.dong,ben.rombaut,ruochen.deng1,justina.lin,arthur.leung1,boyuan.chen1\}@huawei.com}} \\
    \scalebox{0.75}{\texttt{shaowei.wang@umanitoba.ca, dayilin@cs.queensu.ca, ahmed@cs.queensu.ca}}
  }
}

\newcommand{\ourtool}{\textbf{Ours}\xspace}

\begin{document}

\maketitle

\begin{abstract}

LLMs have substantially improved software engineering yet real-world development requires architectural understanding. Such understanding is prohibitively expensive to label manually and impossible to verify through tests alone. We propose an agentic judging pipeline using a strong LLM as a scalable proxy for expert architectural evaluation, comprising two judges: the Architecture Complexity Judge (ACJ), which estimates codebase-specific architectural understanding a task demands, and the Architecture Quality Judge (AQJ), which evaluates patch conformance to repository-specific architectural conventions via source-grounded rubrics. Fine-tuning Qwen3-8B/14B/32B on 3,360 curated instances achieves resolved rates of up to 27.2\% on SWE-bench Verified — up to 540\% over the base model and 256\% over unfiltered fine-tuning. Meanwhile, the trained models achieve strong cross-language generalization and consistent improvements in architectural patch quality.
\end{abstract}

\section{Introduction}\label{sec:intro}



\begingroup
\renewcommand\thefootnote{}\footnote{We provide code and data at \texttt{\url{https://huggingface.co/datasets/spdataset/high_quality_trajectories} and \url{https://anonymous.4open.science/r/full_evaluation_pipeline-E239/README.md}}}
\addtocounter{footnote}{-1}
\endgroup

Recent advances in LLMs have substantially improved software engineering tasks such as code generation~\cite{chen2021evaluating} and bug fixing~\cite{yang2025lingxi,jimenez2024swe}. However, many real-world tasks (e.g., feature implementation, large-scale refactoring, and issue resolution in complex repositories) require architectural understanding: reasoning about module boundaries, dependency structures, and cross-module change propagation~\cite{wan2023software}. Without this capability, LLM-generated solutions are often locally correct but globally harmful, introducing regressions, violating design conventions, or weakening modularity~\cite{liu2024refining}.

Unlike functional correctness, which can be verified through tests or execution-based oracles, architectural quality is inherently subjective.  Properties such as modularity, cohesion, and layering depend on engineering context and expert judgment rather than a single verifiable ground truth. No individual metric such as lines of code or coupling, adequately captures architectural quality, as it emerges from the interaction of multiple structural factors. 
This lack of objective verifiability creates a severe labeling bottleneck. Evaluating architectural quality requires experienced engineers capable of assessing long-range design coherence and dependency management. Such expertise is costly, difficult to scale, and often yields inconsistent annotations due to the subjective nature of architectural judgment and the absence of universally accepted criteria across projects and domains.

To address this challenge, we propose an agentic judging pipeline that uses a strong LLM as a scalable proxy for expert architectural evaluation, enabling automated construction of high-quality SFT data for architectural reasoning. Our pipeline consists of two complementary judges. The \textit{Architecture Complexity Judge} (ACJ) estimates how much codebase-specific architectural understanding a task demands — distinguishing problems that require deep structural reasoning about component relationships, hidden invariants, and cross-boundary effects from those solvable with general programming knowledge alone. The \textit{Architecture Quality Judge} (AQJ) assesses whether a candidate patch conforms to the target repository's architectural conventions, generating a repository- and issue-specific rubric from source evidence before evaluating any patch against it. Both judges operate exclusively through static structural analysis, without executing code or relying on test outcomes, making them applicable to any repository regardless of test availability.

With 3,360 architecturally curated bug fixing instances with trajactory selected by using our pipeline, we fine-tunes (SFT) Qwen3-8B/14B/32B models. On SWE-bench Verified, we achieve resolved rates of 17.4\% - 27.2\%, representing up to 540\% improvement over the base model (without SFT), and 256\% over unfiltered full-data fine-tuning. On SWE-bench Multilingual, despite training exclusively on Python, our approach improves over the base model by 286\%-424\%, demonstrating strong cross-language generalization. Further, our approach increases the proportion of architecturally conformant patches from 61--72\% to 84--94\% across model sizes. Lastly, our approach consistently outperforms metric-based filtering using static metrics (e.g., change in LoC and file size), confirming that agentic architectural judges capture reasoning signals that static metrics cannot replicate.

\section{Background and Related Work}\label{sec:background}

\noindent\textbf{Software Architecture Understanding.}
Beyond functional correctness, real-world software maintenance requires reasoning about system architecture, including module boundaries, layering constraints, and cross-component dependencies. Prior work in software engineering has studied architectural quality through static analysis metrics such as coupling, cohesion, and complexity~\cite{chidamber1994metrics}. While useful for approximate assessment, these metrics are limited in their ability to capture holistic architectural intent, which is often context-dependent and emergent from interactions among multiple design decisions.
Existing metrics are both (i) repository-agnostic and (ii) locally defined~\cite{silva_arch_quality_metrics}, meaning they fail to capture system-specific design conventions such as extension mechanisms, implicit layering rules, or domain-specific module responsibilities. Consequently, two patches with identical metric profiles may differ substantially in architectural validity. 


\noindent\textbf{LLM-as-a-Judge and Rubric-Based Evaluation.}
Recent work has explored using LLMs as evaluators for subjective or hard-to-verify tasks, including summarization quality, reasoning correctness, and preference judgment~\cite{zheng2023judging,wei2024systematic,thakur2025judging}. A key development in this line of research is rubric-guided evaluation, where LLMs generate or follow structured scoring criteria to improve consistency and alignment with human judgments~\cite{hashemi2024llm,rao2026autorubric}. Rubric-based methods have been shown to improve evaluation reliability by externalizing implicit evaluation criteria into explicit, reusable structures, enabling more calibrated and interpretable judgments across diverse evaluation settings~\cite{hashemi2024llm,rao2026autorubric}.

However, existing rubric-based approaches are predominantly task-level and static: they assume a fixed evaluation schema shared across all inputs. This assumption breaks down in software engineering settings, where architectural constraints are inherently repository-specific and evolve across projects. As a result, fixed rubrics risk either over-constraining (penalizing valid project-specific designs) or under-specifying (failing to detect structural violations).




\section{Methodology}\label{sec:method}





\begin{figure*}[t]
    \centering
    \includegraphics[width=\textwidth]{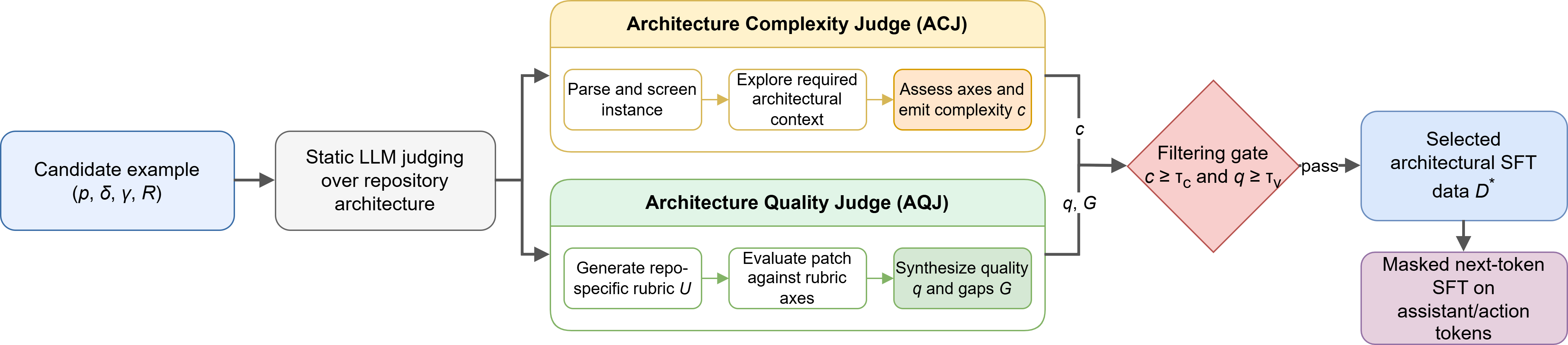}
    \caption{Overview of our data construction workflow. Each candidate example consists of a problem statement $p$, candidate patch $\delta$, reasoning trajectory $\gamma$, and repository $\mathcal{R}$. The Architecture Complexity Judge (ACJ) produces an architectural complexity verdict $c$, while the Architecture Quality Judge (AQJ) produces an architectural conformance verdict $q$ and gap list $\mathcal{G}$. Examples satisfying the complexity threshold $\tau_c$ and quality threshold $\tau_v$ are retained as the selected supervised fine-tuning dataset $\mathcal{D}^{*}$.}
    \label{fig:method-flow}
\end{figure*}

We aim to construct supervised fine-tuning (SFT) data that improves architectural reasoning in LLMs beyond local functional correctness. Given a pool of candidate examples, with each consisting of a problem statement, repository, candidate patch, and LLM-generated reasoning trajectory, our central hypothesis is that an example is valuable for architectural SFT when two conditions hold simultaneously: the training set covers diverse levels of architectural complexity, and the candidate patch conforms well to the target repository's architecture.

Therefore, the judging pipeline has two complementary components (Figure ~\ref{fig:method-flow}). The Architecture Complexity Judge (ACJ) estimates how much repository-specific architectural understanding the task demands. The Architecture Quality Judge (AQJ) first generates a repository- and issue-specific architectural rubric capturing the structural conformance criteria for the affected codebase, then evaluates the candidate patch against that rubric. Examples passing configurable thresholds on both judges are selected for SFT.


\subsection{Architecture Complexity Judge (ACJ)}\label{sec:ACJ}



ACJ measures the architectural complexity of a task: how much codebase-specific architectural understanding a skilled engineer, new to the repository, would need in order to arrive at the patch. This is distinct from general task difficulty. General programming knowledge, external library APIs, domain knowledge, algorithmic difficulty, and implementation effort are outside the scope of ACJ unless they depend on the repository's own architecture. More specifically, ACJ takes a triple as the input and output final complexity verdict 

$$c = ACJ(p, \delta, \mathcal{R})$$

where $p \in \mathcal{P}$ is the natural-language problem statement, $\delta \in \Delta$ is the patch represented as a set of file-level diffs, and $\mathcal{R}$ is the repository codebase providing the architectural context required for evaluation. $c \in \mathcal{C} = \{\texttt{Trivial}, \texttt{Low}, \texttt{Moderate}, \texttt{High}, \texttt{Expert}\}$ is the final complexity verdict.

The complexity judge follows a four-stage procedure:

\noindent\textbf{Step 1: Instance Parsing.} The judge reads the problem statement and patches in parallel, meanwhile extracting: (i) the reported symptom and affected components, (ii) the set of changed files and nature of each change, and (iii) whether the patch bundles multiple independent fixes. Multi-change patches are flagged for per-change evaluation followed by axis-wise aggregation.

\noindent{\textbf{Step 2: Triviality Screening.}} Before full exploration, the judge applies a conservative fast-path: if and only if the patch touches exclusively comments, documentation, string literals, or formatting, with zero logic changes, the instance is immediately labeled trivial and skipped. Any logic modification, however minimal, proceeds to full exploration. This gate prevents premature labeling of small patches as low-complexity.

\noindent{\textbf{Step 3: Patch-Guided Outward Exploration.}} 
A cold-start engineer approaching the problem never sees the patch — they must navigate from the problem description to a correct solution entirely through their understanding of the codebase's structure. We simulate the repository exploration process a cold-start engineer must perform to acquire the codebase-specific architectural knowledge required to produce the patch. Rather than reading the diff as an answer, we use the patch as the starting point to reverse-engineer the architecture knowledge acquired for the patch.
Starting from the changed files, the judge explores the repository by tracing callers, callees, interfaces, data-flow paths, shared state, sibling implementations, and architectural boundaries. Exploration terminates when crossing a clear architectural boundary or when the relevant context is established. For each file read, the judge assesses comprehension density — how hard it is to extract the relevant structural insight from that file, independent of knowing which file to open. Note that comprehension density captures this per-file extractability cost, enabling the final verdict to reflect not just how broadly the engineer must explore but how hard the exploration will be once they know where to look (in Stage 4).  The judge also identifies up to 3 rejected fix paths: plausible-but-wrong approaches a cold-start engineer might attempt without sufficient repo-specific architectural context, and explains why codebase-specific architecture makes each fail. These paths provide falsifiable evidence that the identified structural knowledge is genuinely necessary, not merely incidently observed. A plausible-but-wrong alternative approach and a precise explanation of why codebase-specific architecture makes it fail, demonstrate that the correct solution is not derivable from general programming knowledge alone, and that the identified structural context is what separates a correct solution from a reasonable but incorrect one.

\noindent\textbf{Step 4: Multi-Axis Assessment and Verdict.} Architectural complexity is not a single dimension. A problem can require reading many files but each yielding obvious insights, or require reading only one file whose key invariant is deeply hidden. A single holistic score conflates these distinct costs and loses the discriminative signal needed for dataset stratification. We therefore decompose complexity into five independent axes: 1) Scope of Context; 2) Dependency
Chain Depth, Implicit Knowledge, Coordination Complexity, and Insight Density (more details in Table~\ref{tab:complexity-axes} of Appendix). Each axis targets a different structural dimension before deriving an aggregate verdict. 
Each axis is assessed independently. Finally, ACJ aggregates the resultant five axes into one of five labels (verdict) $\mathcal{C}$. More details can be found in Appedix~\ref{append:ACJ}.


\subsection{Architecture Quality Judge (AQJ)}
AQJ assesses the quality of a generated patch and measures how well it respects the architecture, such as the patterns, boundaries, and structural conventions of the target repo. Architectural quality is inherently repository-specific. There is no universal structural rubric: a layered architecture, an event-driven system, and a plugin-based framework each demand different conformance criteria, and the relevant criteria vary further with issue type. This specificity makes a two-stage design necessary: the repo-specific rubric must be generated from the repository before evaluation can occur. 

\subsubsection{Rubric Generation}

AQJ uses codebase-derived rubrics rather than a fixed set of universal architectural axes. Pre-defined fixed-axis rubrics risked imposing the evaluator's prior assumptions on every repository. The current design asks the judge to infer which architectural properties matter in the given repository and issue context, then commit those properties into a reusable rubric before patch evaluation. The rubric generator takes as input a problem statement $p$ and a repository $\mathcal{R}$, and produces a repository-specific rubric $\mathcal{U}$ consisting of a set of priority-ordered axes, each with prose conformance anchors grounded in the architectural patterns and boundaries found in $\mathcal{R}$:

$$ \mathcal{U} = \text{RG}(p, \delta, \mathcal{R})$$

The rubric generation follows four steps. 

\noindent\textbf{Step 1: Issue Parsing and Issue Type Classification.} 
Issue type determines what structural properties are load-bearing for the rubric: a bug fix calls for pattern preservation and boundary respect, while a feature addition calls for extensibility along the existing extension mechanism. Without this classification, the rubric applies uniform criteria regardless of what kind of change is demanded, producing anchors that are either too strict or too lenient. This step reads the issue report and simultaneously scans the repository's top-level structure, configuration files, and entry points. From the issue report, it extracts the problem description, affected components, expected behavior, and any design clues. Critically, it classifies the issue into one of four types: bug fix, feature addition, refactoring, or architectural change. These categories are grounded in established software maintenance theory — bug fix and architectural change correspond to corrective and adaptive maintenance~\cite{lientz1980software}, refactoring follows the standard definition of behavior-preserving restructuring~\cite{fowler2018refactoring}, and feature addition captures perfective change aimed at extending system capability. 


\noindent\textbf{Step 2: Evidence-Based Architecture Exploration.}
This step explores the repository to identify components and their responsibilities, dependency direction and layering, architectural styles, design patterns with location and scope, cross-cutting conventions, and critical boundaries with their enforcement mechanisms. Every claim must cite concrete evidence (e.g., specific files, imports, or directory structure) and receive a confidence score (i.e., 0 - 1). We only keep the patterns with confidence $\geq 0.6$. This evidence requirement guards against the primary failure mode of generic rubrics: axes like ``follows design patterns'' or ``respects layering'' are abstract enough to match any patch, yielding labels with no discriminative signal. Grounding axes in what is actually structurally present ensures the rubric captures the architectural commitments this codebase has made, not a template of what good architecture should look like in general.

\noindent\textbf{Step 3: Axis Derivation and Priority Assignment.}
From the exploration results, the judge derives rubric axes by identifying which structural properties are most significant for the issue type and most at risk from the likely patch, assigning each a priority: primary (load-bearing — a violation constitutes a fundamental structural problem), secondary (important but not decisive alone), or minor (worth noting but rarely decisive). Not all structural properties are equally consequential. For example, treating a naming convention violation with the same weight as a layer boundary crossing produces rubrics where peripheral concerns drown out what actually matters. The final assessment is anchored to primary axes, with secondary and minor axes providing nuance rather than driving the outcome. 

\noindent\textbf{Step 4:  Prose Anchor Generation.} For each axis, the judge produces two prose anchors: an \textit{ideal} conformance anchor describing what a patch that fully respects this structural dimension looks like, and a \textit{poor} conformance anchor describing the structural problems of a non-conformant patch. Anchors must describe structural properties and invariants, rather than specific implementation mechanisms (e.g., ``The fix adds SQLiteNumericMixin to Window's inheritance chain'').

\subsubsection{Patch Evaluation}

The patch evaluator takes as input a problem statement $p$, a patch $\delta$, the generated rubric $\mathcal{U}$, and the repository $\mathcal{R}$, and produces a categorical architectural conformance verdict $q \in Q = \{\texttt{High}, \texttt{High\_with\_concerns}, \texttt{Acceptable}, \texttt{Low}\}$, where High (no primary-axis violation, empty gap list), High\_with\_concerns (no primary-axis violation, non-empty gap list), Acceptable (contained or justified primary-axis deviation), or Low (significant primary-axis violation, pattern bypass, or boundary crossing), together with a set of gaps $\mathcal{G}$ enumerating concrete structural divergences from ideal conformance:
$$(q, \mathcal{G}) = \text{PE}(p, \delta, \mathcal{U}, \mathcal{R})$$

Similar to rubric generation, we conduct issue/patch parsing and codebase exploration to collect architectural context related to the issue and patch. We then follow the three steps for evaluation. 

\noindent\textbf{Step 1: Parallel Per-Axis Evaluation.}
To prevent a pre-formed overall judgment from contaminating individual axis assessments, each rubric axis is evaluated independently by a dedicated sub-agent with all axes dispatched simultaneously. Each sub-agent receives only its own axis definition, both prose anchors, the patch analysis, and the exploration findings, and produces a prose assessment describing where the patch falls on the spectrum between ideal and poor conformance with file-level structural evidence. Parallel dispatch further ensures that no earlier axis assessment can anchor later ones, preserving the independence of the evidence base across all axes.

\noindent\textbf{Step 2: Impact Analysis and Cross-Axis Calibration.} 
For each primary axis, the judge analyzes structural effects beyond the patch's direct file scope (e.g., subclasses of modified base classes, callers of modified methods, and modules importing changed symbols), and assesses whether the patch introduces unintended structural coupling or breaks downstream invariants. This is necessary because a change that appears locally conformant on every individual axis may nevertheless propagate structural effects only visible in downstream consumers, a blast radius invisible to per-axis independent analysis. The judge then reviews all axis assessments together for internal consistency, identifying and adjudicating inconsistent severity language, contradictory claims, and mismatches between axis language and the intended verdict. The calibration step resolves inconsistencies across independently assessed axes, ensuring the overall assessment is internally coherent and its reasoning auditable.

\noindent\textbf{Step 3: Verdict Synthesis.}
Lastly, we synthesize all axis assessments and impact analysis into a four-tier categorical verdict anchored to primary axes: High, High\_with\_concerns, Acceptable, or Low. Anchoring to primary axes prevents secondary concerns from overriding fundamental structural violations and minor-axis strengths from masking them. Regardless of the verdict, an explicit gap list enumerates every concrete structural divergence from ideal conformance — ensuring that even strong verdicts carry actionable structural feedback for improving architectural reasoning.

More details of the prompts can be found in Appendix~\ref{append:ACJ}

\subsection{Filtering Thresholds and Training}

Let $\mathcal{D} = {x_i}$ denote the pool of candidate examples, where each example is a triple:

$$x = (p, \delta, \gamma, \mathcal{R})$$

,with $p$ the problem statement, $\delta$ the candidate patch, $\gamma$ the trajectory, and $\mathcal{R}$ the repository. 
Given Architecture Complexity Judge (ACJ) threshold $\tau_c \in \mathcal{C}$ and Architecture Quality Judge (AQJ) threshold $\tau_v \in \mathcal{V}$, the selected dataset is:

$$\mathcal{D}^* = \{ x \in \mathcal{D} \mid c \geq \tau_c \ \wedge \ q \geq \tau_v \}$$

where $c$ and $v$ are the verdicts generated by ACJ and AQJ for $x$. 

For training, we optimize a masked next-token cross-entropy loss over the filtered trajectory dataset $\mathcal{D}^{*}$. Specifically, only assistant-generated trajectory tokens are supervised, while user messages and environment observations are masked out from the loss.


\section{Experimental Settings}\label{sec:experimentalsetting}

\noindent\textbf{\underline{Models}} 
To verify both the effectiveness of our approach, we conduct experiments open-source models with different size Qwen3-8B, Qwen3-14B, and Qwen3-32B~\cite{yang2025qwen3}.

\noindent\textbf{\underline{Dataset and metrics}}
Our initial training data is derived from a repository-level software engineering dataset constructed from real-world GitHub repositories using the RepoForge environment construction pipeline~\cite{chen2025repoforge}. The original dataset consists of 5,000 high-quality Python trajectories generated using the MiniMax-M2.5 model~\cite{minimax_m25_2026} and selected via teacher pass-rate-based filtering. We use ACJ $\geq$ Trivial and AQJ $\geq$ Acceptable as our threshold to select data from the training dataset, and ended up with 3,360 architecturally curated instances. We use trivial as the threshold as we would like to include samples with diverse complexity of architecture understanding. 

We evaluate our approach on two benchmarks: \textbf{SWE-bench-verified} ~\cite{jimenez2024swe}, which contains 500 verified bug-fixing instances, and \textbf{SWE-bench Multilingual}~\cite{yang2025swesmith}, which contains 300 bug-fixing instances in 9 non-Python languages (e.g., Ruby and Java). 
We use SWE-bench Multilingual to test the model's cross-language generalization after training with our selected data.  Note that there is no overlap between the training and evaluation datasets.
In all experiments, we use OpenHands \citep{wang2025openhands} as the scaffold, with max 300 turns per instance, and report performance using the \textbf{resolved rate} under a Pass@1 setting, where a task is considered successfully resolved if the model-generated patch passes all unit tests specified on the first attempt~\cite{traeresearchteam2025traeagent,yang2025lingxi,jimenez2024swe}. 

\noindent\textbf{\underline{Baselines}} 
To evaluate the effectiveness of our approach, we compare it against the following baselines: \textbf{Original}, the base model without any fine-tuning; \textbf{FullData}, a model fine-tuned on the entire training dataset; and \textbf{\ourtool}, a model fine-tuned using the subset of data selected by our approach.

\noindent\textbf{\underline{Training and implementation}}
In our experiments, we fine-tune all models using AdamW ($\beta_1=0.9$, $\beta_2=0.95$) with BF16 precision. We use a cosine learning rate schedule with an initial learning rate of $5\times10^{-6}$, a minimum learning rate of $1.25\times10^{-7}$, and a 1\% warmup ratio. We apply gradient clipping with a norm of 1.0 and a weight decay of 0.1. All models are trained for 3 epochs. 
The global batch size is set to 32 for Qwen3-8B, and 16 for Qwen3-14B and Qwen3-32B. 
Following previous work on agent-based software engineering evaluation, we use \textbf{OpenHands}~\cite{wang2025openhands} as the execution scaffold to enable repository-level interaction, patch application, and automated testing. 

\section{Results}\label{sec:result}

\subsection{RQ1: Effectiveness of selected data in improving models}

\begin{table*}
\centering
\small
\begin{tabular}{l|l|l|l|l|l|l|l|l}
\hline
\textbf{Model} &
\textbf{Data} &
\textbf{Patches} &
\textbf{Resolved} &
\textbf{Precision} &
\textbf{High (\%)} &
\textbf{HWC (\%)} &
\textbf{Acc. (\%)} &
\textbf{Low (\%)} \\
\hline

\multirow{3}{*}{Qwen3-8B}
& Original   & 228   & 3.2\%  & 5.7\%    & 10.5  & 50.4  & 17,1  & 21.9  \\ \cline{2-9}
& FullData   & 216   &    6.8\%   & 15.7\%   & +10.8 & +15.8 & -10.2 & -16.3 \\ \cline{2-9}
& Ours       & 262   & \textbf{17.4\%} & \textbf{30.1\%}   & +4.0  & +18.7 & -5.3  & -17.3 \\
\hline

\multirow{3}{*}{Qwen3-14B}
& Original   & 200 & 4.6\%  & 11.0\%    & 12.5  & 51.5  & 14.0  & 22    \\ \cline{2-9}
& FullData   & 236 & 9.4\%  & 20.0\%    & +1.9  & +18.0 & -5.1  & -14.8 \\ \cline{2-9}
& Ours       & 186 & \textbf{17.8\% } & \textbf{47.3\%}    & +2.0  & +21.1 & -4.3  & -18.8 \\
\hline

\multirow{3}{*}{Qwen3-32B}
& Original   & 193 & 9.0\%  & 20.2\%    & 7.3   & 64.8  & 9.8   & 18.1  \\ \cline{2-9}
& FullData   & 309 & 15.6\%  & 25.0\%    & +9.9  & +11.3 & -5.3  & -15.8 \\ \cline{2-9}
& Ours       & 274  & \textbf{27.2\% } &\textbf{49.53\%}   & +9.9  & +11.5 & -5.4  & -15.9 \\
\hline

\end{tabular}
\caption{Comparison of our approach with baselines (i.e., Original and FullData) across the studied models on SWE-bench Verified in terms of pass@1 resolve rate (\textit{Resolved}), architecture quality distribution of all patches (measured by AQJ: \textit{High}, \textit{HWC (High with concerns)}, \textit{Acc. (Acceptable)}, and \textit{Low}), and \textit{precision} (measured as the portion of generated patches that successfully resolved the issue over total number of generated patches (\textit{Patches})), Note that AQJ scores are shown as relative change to the respective model baseline, while the baseline shows the absolute proportion of total patches per AQJ label.}
\label{tab:aqj_scores}
\end{table*}

Table~\ref{tab:aqj_scores} shows that \textbf{our approach consistently outperforms both baselines across all model sizes on SWE-bench. For instance, our approach improves the resolved rate by 174\% - 256\%, compared to using the full dataset.} Compared to the Original baseline, our approach improves the resolved rate from 3.2\% to 17.4\% on Qwen3-8B (540\% improvement), from 4.6\% to 17.8\% on Qwen3-14B (387\% improvement), and from 9.0\% to 27.2\% on Qwen3-32B (302\% improvement). Our approach also substantially outperforms the FullData baseline across all model sizes, improving from 6.8\% to 17.4\% on Qwen3-8B (256\% improvement), from 9.4\% to 17.8\% on Qwen3-14B (189\% improvement), and from 15.6\% to 27.2\% on Qwen3-32B (174\% improvement), demonstrating that selecting a smaller but architecturally meaningful subset is strictly more effective than training on all available data indiscriminately.


The AQJ quality distribution in Table~\ref{tab:aqj_scores} further reveals that \textbf{our approach consistently shifts patch quality toward higher architectural conformance compared to the Original baseline}. Across all model sizes, our approach substantially increases the proportion of High and High-with-concerns (HWC) patches while reducing Low-conformance patches. For Qwen3-8B, the combined High+HWC proportion increases from 60.9\% to 83.6\%; for Qwen3-14B, from 64.0\% to 87.1\%; and for Qwen3-32B, from 72.1\% to 93.5\%. This confirms that fine-tuning on architecturally curated data not only improves task resolution but also steers the model toward generating patches that better respect the structural conventions of the target repository. \textbf{Compared to FullData, our approach achieves comparable architectural quality} — with High+HWC proportions of 83.6\% vs. 87.5\% on Qwen3-8B, 87.1\% vs. 83.9\% on Qwen3-14B, and 93.5\% vs. 93.3\% on Qwen3-32B — despite using a significantly smaller and more selective training set. For instance, we show an example of high and low quality patches in Figure ~\ref{fig:patch_compare} in Appendix ~\ref{app:patch_compare}, where the high quality patch directly addresses the root cause of the issue while the low quality patch does not and introduces extra fragility.

Lastly, our approach shows a 293\% to 428\% improvement over baseline and 91.7\% to 136\% over FullData in the generated patch precision across all model sizes, meaning the model is much more likely to generate a successful, issue-resolving patch. Improving the model's patch precision can lead to downstream efficiency gains by requiring fewer attempts per successful fix, and consequently, lowering compute costs and improving user trust in the system.



\subsection{RQ2: Cross-language generalization}

\begin{table}
\centering
\small
\begin{tabular}{l|l|l}
\hline
\textbf{Model} &
\textbf{Data} &
\textbf{SWE-Multi} \\
\hline

\multirow{3}{*}{Qwen3-8B}
& Original   & 1.4\%   \\ \cline{2-3}
& FullData   & 3.0\%  \\ \cline{2-3}
& Ours       & \textbf{4.0\% } \\
\hline

\multirow{3}{*}{Qwen3-14B}
& Original   & 4.3\%  \\ \cline{2-3}
& FullData   & 9.3\%  \\ \cline{2-3}
& Ours       & \textbf{9.7\% } \\
\hline

\multirow{3}{*}{Qwen3-32B}
& Original   & 3.3\%   \\ \cline{2-3}
& FullData   & 11.7\%   \\ \cline{2-3}
& Ours       & \textbf{14\% } \\
\hline

\end{tabular}
\caption{Comparison of our approach with baselines across the studied models on SWE-bench Multilingual in terms of resolved rate.}
\vspace{-0.2in}
\label{tab:results_multi}
\end{table}

\textbf{Our approach generalizes to other programming languages despite being trained exclusively on Python instances} as shown in Table~\ref{tab:results_multi}. Evaluated SWE-bench Multilingual, which spans 9 languages beyond Python, our approach improves over the Original baseline by 286\% - 420\% across different models. Similarly, our approach outperforms FullData consistently across all model sizes. These consistent gains suggest that the architectural reasoning patterns learned from Python training data can effectively transfer to other languages, indicating that architectural understanding is to a meaningful degree language-agnostic — grounded in structural properties such as module boundaries, dependency direction, and abstraction patterns rather than language-specific syntax.

\subsection{RQ3: Collinearity with static metrics}
\label{sec:collinearity}

Given the token and resource costs associated with running an agentic judge, we investigated whether it could be replicated using several established static software architecture quality metrics. 
Following previous studies~\cite{silva_arch_quality_metrics}, we implemented a subset of 30 unique architecture quality metric groups measured at file, package/component, and repository level.
Since our training data is based on evaluating the quality of generated patches, we capture the impact of applying the patch by measuring the relevant metrics as the delta between metric scores before and after the patch is applied. 
Additionally, we implemented seven groups of patch-specific metrics, relating to lines-of-code (LoC), number of affected files, structural changes, and others. Overall, this yields 157 deterministic metrics calculated for each sample in our training data. 
We calculate the Spearman correlation between the collected static metrics and the ACJ and AQJ scores generated by our agentic judges. See Appendix~\ref{app:metrics} for a detailed listing of metrics and their descriptions.


\textbf{Our agentic-based ACJ/AQJ scores cannot be replicated using static code quality and complexity metrics.}
For AQJ scores, no single static metric strongly correlates with the judge scores; the top-ranked features are primarily change-size metrics (e.g., file size/LoC delta), with the highest Spearman correlation of only $\rho=-0.24$. Similarly, for ACJ scores, the most correlated features are also dominated by change-size metrics (e.g., file size/LoC delta), with the highest observed Spearman correlation being $\rho=0.32$. Overall, while AQJ and ACJ share a broadly similar set of moderately correlated features, no individual static metric is strongly predictive of the agentic judges' assessments. Additional analysis is in Appendix \ref{app:metrics}.

We also investigated whether it is possible to replicate the agentic filtering pipeline using static metrics. 
As a proof-of-concept, we selected a subset of four highly-ranked metrics from the top-20 ones described in our collinearity analysis. 
We then threshold these scores into four-bin regions identical to the labels used by ACJ and AQJ, respectively.
The final class is then decided by averaging the four scores and quantile-binning into the final class (Appendix \ref{app:metrics}). 
Finally, we used these metrics to assign ACJ/AQJ-style scores to training data samples in Section ~\ref{sec:experimentalsetting} and then finetune a Qwen3-8B model. We use the same filtering threshold as RQ1 on the training dataset (i.e., ACJ $\geq$ Trivial and AQJ $\geq$ Acceptable). 
Lastly, we compared the performance of this model with an identically filtered one using our method (Table \ref{tab:metric_threshold_comparison}).

\begin{table}[h]
\centering
\footnotesize
\setlength{\tabcolsep}{5pt}
\renewcommand{\arraystretch}{1.2}
\begin{tabular}{l|c|c|c}
\hline
\textbf{Filter} & \textbf{Size} & \textbf{SWE-Verified} & \textbf{SWE-Multi} \\
\hline
Original       & 5,000  & 6.8\% (34) & 5.0\% (15) \\ \cline{1-4}
Metrics   & 1,548 & 14.4\% (72) & 4.7\% (14) \\ \cline{1-4}
\ourtool     & 3,360 & \textbf{17.4\% (87)} & \textbf{6.7\% (20)} \\ \cline{1-4}

\hline
\end{tabular}
\caption{Comparison of using \ourtool and metric-based filtering on Qwen3-8B.}
\label{tab:metric_threshold_comparison}
\end{table}

\textbf{Our approach outperforms metric-based filtering.} Table~\ref{tab:metric_threshold_comparison} shows that filtering by static software engineering metrics improves the resolved rate from 6.8\% to 14.4\% on SWE-bench Verified, but degrades performance from 5.0\% to 4.7\% on SWE-bench- Multilingual, suggesting that metric-based thresholds fail to generalize across evaluation settings. Our approach consistently improves over both baselines on both benchmarks, demonstrating more robust data selection. This result, together with the collinearity analysis, indicates that our LLM-based judges capture semantic architectural reasoning that deterministic metrics cannot fully express, leading to higher-quality filtering and stronger downstream model performance.



\section{Discussion}\label{sec:discussion}

\subsection{Can we predict the agentic-based ASJ/ACJ score by using static metrics?}
Our collienarity analysis showed that classical software architecture quality metrics individually have low correlation with our agentic scores and cannot match the performance with basic thresholding; however, they do come close when used together in combination with each other. To investigate further, we used the entirety of 157 metrics as features to train random forest (RF) models to predict ACJ/AQJ multiclass scores (same training data as Section ~\ref{sec:collinearity}, 80-20\% train-test split). While the ACJ model only reached 49\% PR AUC, the AQJ model reached an impressive 75\% PR AUC. This points to the fact that there is joint, non-linear information shared across static metrics that is informative for AQJ scores. However, this is not the case for ACJ scores. One possible explanation is that complexity is both task- and process-dependent: arriving at an accurate estimate requires the agent to actively explore the repository, model its architecture and design choices, and reason about how those choices constrain the solution path. Nuanced signals such as documentation specificity, implicit conventions, and cross-boundary dependencies cannot be captured by static code structure metrics.


\section{Conclusion}
We presented an agentic judging pipeline for constructing architecturally curated SFT data, addressing the labeling bottleneck that makes architectural quality difficult to assess at scale. Our two complementary judges, ACJ for task complexity and AQJ for patch conformance, operate through static structural analysis without code execution, producing training signal that static metrics cannot replicate. Fine-tuning on our curated data consistently outperforms both untuned and fully fine-tuned baselines across model sizes, and achieves strong cross-language generalization. Our approach also improves architectural patch conformance alongside task resolution. These results suggest that architectural reasoning is a learnable and transferable capability when training data is selected with explicit structural quality criteria.

\section*{Limitations}
In this paper, we evaluated our filtering strategy on three open source models and two widely used benchmarks. Our findings might be not generalized to other models, we encourage future research to measure the effective of our approach on more models.



\bibliography{custom}

\appendix

\section{Appendix}

\label{sec:appendix}

\subsection{Implementation details of Skills}\label{append:ACJ}
\begin{strip}
\begin{tcolorbox}[
    title=Prompt for Evidence-Based Architecture Exploration (ACJ),
    colback=gray!5,
    colframe=black
]
\footnotesize

You are evaluating how much architectural understanding of a codebase is required to solve a specific problem. You are NOT evaluating the quality of the solution --- you are assessing how much of the system's architecture someone would need to understand before they could arrive at the correct solution. \\

IMPORTANT SCOPE CONSTRAINT: Only count knowledge that is specific to how this codebase is structured. Do NOT count: \\
- General language semantics (e.g., Python GIL behavior, reference counting) \\
- External library or framework APIs (e.g., how fsspec works, what SQLAlchemy does) \\
- Domain knowledge (e.g., HTTP spec, OAuth protocol) \\
- Algorithmic complexity \\

\textbf{Instance} \\

Problem statement: \\
\{paste\_full\_problem\_statement\} \\

Patch (files changed): \\
\{paste\_list\_of\_changed\_files\_and\_summaries\} \\

Codebase location: \\
\{repo\_path\} \\

\textbf{Exploration Instructions} \\

Starting from each file changed in the patch, explore outward to understand the architectural context. For each changed file: \\

1. \textbf{Immediate context}: Read the changed file. What module/component does it belong to? What is its role? \\

2. \textbf{Callers and callees}: What calls the changed code? What does the changed code call? Trace outward until you cross an architectural boundary or the relevant context is clear --- do not trace further than needed. \\

3. \textbf{Cross-boundary context} (pursue only when step 2 reveals cross-boundary dependencies): Investigate whichever of the following are relevant to this specific change --- skip any that are not: \\
- Interfaces, contracts, or protocols the changed code participates in \\
- Data flow paths into and out of this code \\
- Shared state or invariants spanning beyond the changed code (codebase-specific only) \\
- Architectural patterns governing this area (layering, plugin systems, event-driven, etc.) \\
- Sibling implementations that establish conventions a solver would need to follow \\

4. \textbf{Comprehension density}: For each file you read, assess: if a cold-start engineer already knew to open this file, how hard would it be to extract the relevant architectural insight from it? Is the key invariant documented, visible in the code, or only apparent from tracing runtime behavior? \\

5. \textbf{Rejected fix paths}: Identify 1--3 plausible-but-wrong fix approaches that a cold-start engineer might attempt and explain why each would fail due to codebase-specific architecture. If no such paths exist --- i.e., the correct fix is self-evident from the changed file alone without codebase-specific knowledge --- state that explicitly (this is evidence for a lower verdict). \\

\textbf{Output} \\

Provide a single combined summary (do not repeat findings per-file and then again in a summary): \\

- \textbf{Files explored}: Enumerated list of all distinct file paths read (one per line, each listed once) \\
- \textbf{Areas/modules}: Total count of distinct areas explored \\
- \textbf{Architectural findings}: Key codebase-specific relationships, invariants, and conventions discovered. For each, note whether it is visible in the file itself, discoverable from sibling code, or only emergent from cross-file interaction. \\
- \textbf{Coordination assessment}: Whether the changes themselves vs the reasoning to arrive at them require cross-boundary coordination (note when these diverge) \\
- \textbf{Comprehension density}: How hard is it to extract the relevant architectural understanding once you know where to look? \\
- \textbf{Rejected fix paths}: Each rejected path and why it fails due to codebase-specific architecture \\
\end{tcolorbox}
\begin{tcolorbox}[
    title=Prompt for Architectural Rubric Generation (AQJ),
    colback=gray!5,
    colframe=black
]
\footnotesize

You are generating an architectural rubric YAML file. Write the file at the path specified below. Do not create any other files. \\

\textbf{Output File} \\
Path: \{\{output\_yaml\_path\}\} \\

\textbf{Issue Context} \\
Issue file: [filename only, e.g. issue-report.md] \\
Issue summary: [one-sentence summary grounded in issue description] \\
Issue type: [bug\_fix | feature\_addition | refactoring | architectural\_change] \\
Issue type rationale: [one sentence explaining the classification] \\
Risk level: [low | medium | high] \\
Affected components: [list] \\
Generated date: [YYYY-MM-DD] \\
Instance id: \{\{instance\_id\}\} \\

\textbf{Axes with Priority} \\
\texttt{[For each axis determined in Step 3, one entry each:]} \\
\texttt{[axis\_name]: priority=[primary | secondary | minor], rationale=[why this priority given what exploration found]} \\

\textbf{Architecture Exploration Results} \\
\texttt{[Paste the full output from the Step 2 exploration sub-agent here]} \\

\textbf{Instructions} \\

Derive axis names and descriptions from what patterns, boundaries, and structural properties the exploration results surfaced. Include only axes for which the exploration produced concrete evidence (confidence >= 0.6). Do not include axes for structural properties that are absent or irrelevant in this codebase. \\

Use the axis priorities provided above. Do not reassign priorities. \\

For each axis, write two prose anchors:
\begin{itemize}
\item \textbf{ideal\_conformance}: A paragraph describing what a patch looks like that fully respects this architectural dimension. Be specific to this codebase — reference actual patterns, files, and conventions found during exploration.
\item \textbf{poor\_conformance}: A paragraph describing what violations of this dimension look like. Be specific — describe the kinds of structural problems that would indicate the patch disregards this aspect of the architecture.
\end{itemize}

These anchors are qualitative guides, not scoring criteria. They describe the ends of a spectrum; a real patch will fall somewhere between them. \\

\textbf{CRITICAL — Property-Based Anchors}: Ideal conformance anchors must describe the structural property or invariant the fix must achieve, not the specific function, method, or code pattern that achieves it. Multiple mechanisms may satisfy the same property. The evaluator should assess whether the property holds, not whether a specific mechanism was used. \\

Do NOT include:
- Numeric scores or levels (no 0-5 scale)
- Numeric weights (no decimal weights summing to 1.0)
- Scoring guides or formulas
- Grade scales or thresholds \\

Write a YAML file with this structure:

\end{tcolorbox}

\begin{tcolorbox}[
    title=Single-Axis Architectural Assessment Prompt,
    colback=gray!5,
    colframe=black
]
\footnotesize

You are evaluating a single axis of an architectural rubric. Your ONLY job is to write a prose assessment of the patch on the [AXIS\_NAME] axis. Do not evaluate other axes. Do not evaluate implementation correctness — only structural conformance. Do not assign any numeric scores. \\

\textbf{SCAFFOLDING FILE EXCLUSION}: The following files have been classified as scaffolding (debug scripts, reproduce scripts, test files in repo root, etc.) and are EXCLUDED from architectural assessment. Do not consider them when evaluating this axis. They are not part of the production architecture: \\
\texttt{[List scaffolding\_files from Step 1 patch analysis, or ``None identified'' if empty]} \\

\textbf{Issue context} \\
\texttt{[issue\_summary from rubric metadata]} \\
Issue type: [issue\_type] \\

\textbf{Patch summary} \\
\texttt{[paste patch analysis from Step 1]} \\

\textbf{Codebase exploration findings} \\
\texttt{[paste full exploration output from Step 2]} \\

\textbf{Axis definition} \\
Name: \texttt{[axis\_name]} \\
Priority: \texttt{[primary | secondary | minor]} \\
Description: \texttt{[axis description from rubric]} \\
Baseline context: \texttt{[baseline\_context from rubric]} \\

\textbf{Quality anchors (read BOTH before writing your assessment)} \\

\textit{Ideal conformance:} \\
\texttt{[paste ideal\_conformance from rubric]} \\

\textit{Poor conformance:} \\
\texttt{[paste poor\_conformance from rubric]} \\

\textbf{Assessment instructions} \\

1. Read BOTH quality anchors (ideal and poor) before writing anything. \\
2. Analyze the patch's structural characteristics on this dimension. \\
3. Write a prose assessment that describes: \\
   - What the patch does on this dimension (specific structural changes) \\
   - How the patch's characteristics compare to the ideal anchor \\
   - How the patch's characteristics compare to the poor anchor \\
   - Where on the spectrum between ideal and poor this patch falls \\
   - Any nuances, trade-offs, or context that affect the assessment \\
4. Cite specific evidence for every claim (file:line references). \\
5. Focus on structural conformance ONLY. Ignore bugs, syntax errors, runtime behavior. \\
6. Do NOT include scaffolding files in your assessment — they are out of scope. \\

\textbf{Required output format} \\

Return ONLY the following YAML block. No prose before or after it. \\

\texttt{[axis\_name]:} \\
\quad \texttt{priority: "[primary | secondary | minor]"} \\
\quad \texttt{assessment: |} \\
\quad\quad [Your prose assessment — multiple paragraphs as needed. Describe what the patch does structurally on this dimension, how it compares to the ideal and poor anchors, and where it falls on the spectrum. Be specific and cite evidence.] \\
\quad \texttt{evidence:} \\
\quad\quad - "[file:line — description of structural evidence]" \\
\quad\quad - "[file:line — description of structural evidence]"

\end{tcolorbox}
\end{strip}

\subsection{Example of different quality patches under AQJ scores}
\label{app:patch_compare}
\onecolumn
\begin{center}


\begin{framed}
\begin{verbatim}
diff --git a/monai/transforms/utils.py b/monai/transforms/utils.py
index 11608503..02523cc4 100644
--- a/monai/transforms/utils.py
+++ b/monai/transforms/utils.py
@@ -1713,6 +1713,8 @@ def convert_to_contiguous(
 elif isinstance(data, Mapping):
     return {k: convert_to_contiguous(v, **kwargs) for k, v in data.items()}
 elif isinstance(data, Sequence) and not isinstance(data, bytes):
+        if isinstance(data, tuple):
+            return tuple(convert_to_contiguous(i, **kwargs) for i in data)
     return [convert_to_contiguous(i, **kwargs) for i in data]
 else:
     return data
\end{verbatim}
\end{framed}

\centering
\small (a) AQJ high scoring patch

\vspace{0.8em}


\begin{framed}
\begin{verbatim}
diff --git a/monai/data/utils.py b/monai/data/utils.py
index 14217e91..e1dbd3e0 100644
--- a/monai/data/utils.py
+++ b/monai/data/utils.py
@@ -474,7 +474,7 @@ def collate_meta_tensor(batch):
     if isinstance(elem_0, Mapping):
         return {k: collate_meta_tensor([d[k] for d in batch]) for k in elem_0}
     if isinstance(elem_0, (tuple, list)):
-        return [collate_meta_tensor([d[i] for d in batch]) for i in range(len(elem_0))]
+        return type(elem_0)(collate_meta_tensor([d[i] for d in batch]) for i in range(len(elem_0)))

     return default_collate(batch)
\end{verbatim}
\end{framed}

\centering
\small (b) AQJ low scoring patch

\vspace{0.5em}

\captionof{figure}{
Comparison between AQJ high-scoring and low-scoring patches for issue
\detokenize{Project-MONAI__MONAI-6849}. The root cause of
  this issue is that \texttt{convert\_to\_contiguous} unconditionally converts all
  \texttt{Sequence} types (including \texttt{tuple}) into \texttt{list}. This type change
   propagates downstream: \texttt{list\_data\_collate} treats the resulting list as a
  collection of individual elements to be stacked, causing a shape mismatch error when
  the tensors have different sizes. The high-scoring patch (a) directly addresses the root cause
  by adding an explicit type check for \texttt{tuple} within
  \texttt{convert\_to\_contiguous}, preserving the original type throughout the pipeline.
   This fix is minimal, safe, and consistent with the principle of fixing bugs at their
  source. In contrast, the low-scoring patch (b) modifies
  \texttt{collate\_meta\_tensor}, a downstream function, by dynamically reconstructing
  the container type via \texttt{type(elem\_0)(...)}. While this may coincidentally
  resolve the reported error, it does not address the root cause:
  \texttt{convert\_to\_contiguous} still silently converts tuples to lists, which may
  trigger failures in other downstream consumers. Furthermore, the use of
  \texttt{type(elem\_0)(generator)} introduces fragility, as not all \texttt{Sequence}
  subclasses accept a generator as a constructor argument.
  }

\label{fig:patch_compare}

\end{center}
\twocolumn

\subsection{Studied software architecture quality metrics}
\label{app:metrics}

\begin{table}[ht]
\centering
\small
\begin{tabular}{|l|c|p{9cm}|}
\hline
\textbf{Feature} & \textbf{Spearman $\rho$} & \textbf{Thresholds (High $\rightarrow$ Low)} \\
\hline

\texttt{patch\_total\_nloc\_touched} 
& $-0.138$ 
& $\leq 50$ $\rightarrow$ High; $51$--$200$ $\rightarrow$ HwC; $201$--$500$ $\rightarrow$ Acceptable; $> 500$ $\rightarrow$ Low \\

\hline

\texttt{file\_delta\_complexity\_avg\_cc} 
& $-0.136$ 
& $\leq 0$ $\rightarrow$ High; $(0,1]$ $\rightarrow$ HwC; $(1,5]$ $\rightarrow$ Acceptable; $> 5$ $\rightarrow$ Low \\

\hline

\texttt{file\_delta\_complexity\_avg\_mi} 
& $+0.126$ 
& $\geq 0$ $\rightarrow$ High; $[-2,0)$ $\rightarrow$ HwC; $[-5,-2)$ $\rightarrow$ Acceptable; $< -5$ $\rightarrow$ Low \\

\hline

\texttt{patch\_avg\_shared\_attr\_ratio} 
& $-0.084$ 
& $\leq 0.000$ (p40) $\rightarrow$ High; $\leq 0.886$ (p92) $\rightarrow$ HwC; $\leq 1.000$ (p98) $\rightarrow$ Acceptable; $> 1.000$ $\rightarrow$ Low \\

\hline
\end{tabular}
\caption{Metrics and respective thresholds for ACJ-style labeling.}
\label{tab:acj_thresholds}
\end{table}

\begin{table}[ht]
\centering
\small
\begin{tabular}{|l|c|p{9cm}|}
\hline
\textbf{Feature} & \textbf{Spearman $\rho$} & \textbf{Thresholds (trivial $\rightarrow$ high)} \\
\hline

\texttt{patch\_total\_nloc\_touched}
& $+0.205$
& $\leq 50$ $\rightarrow$ trivial; $51$--$200$ $\rightarrow$ low; $201$--$500$ $\rightarrow$ moderate; $>500$ $\rightarrow$ high \\
\hline

\texttt{patch\_functions\_modified}
& $+0.221$
& $0$--$1$ $\rightarrow$ trivial; $2$--$3$ $\rightarrow$ low; $4$--$5$ $\rightarrow$ moderate; $\geq 6$ $\rightarrow$ high \\
\hline

\texttt{repo\_delta\_size\_loc}
& $+0.323$
& $\leq 0$ $\rightarrow$ trivial; $1$--$10$ $\rightarrow$ low; $11$--$50$ $\rightarrow$ moderate; $>50$ $\rightarrow$ high \\
\hline

\texttt{patch\_avg\_shared\_attr\_ratio}
& $+0.116$
& $\leq 0.000$ (p40) $\rightarrow$ trivial; $\leq 0.577$ (p83) $\rightarrow$ low; $\leq 1.000$ (p100) $\rightarrow$ moderate; $>1.000$ $\rightarrow$ high \\
\hline

\end{tabular}
\caption{Metrics and respective thresholds for AQJ-style labeling.}
\label{tab:aqj_thresholds}
\end{table}

\begin{table}[ht]
\centering
\small
\begin{tabular}{|l|c|p{9cm}|}
\hline
\textbf{Metric} & \textbf{Used In} & \textbf{References} \\
\hline

\texttt{patch\_total\_nloc\_touched}
& ACJ, AQJ
& \cite{smartbear2006}, \cite{google_small_cls} \\
\hline

\texttt{patch\_functions\_modified}
& ACJ
& \cite{rigby2013}, \cite{purushothaman2005} \\
\hline

\texttt{repo\_delta\_size\_loc}
& ACJ
& \cite{smartbear2006}, \cite{google_small_cls} \\
\hline

\texttt{file\_delta\_complexity\_avg\_cc}
& AQJ
& \cite{mccabe1976}, \cite{nist1996} \\
\hline

\texttt{file\_delta\_complexity\_avg\_mi}
& AQJ
& \cite{coleman1994}, \cite{oman1992}, \cite{microsoft_mi} \\
\hline


\end{tabular}
\caption{Supporting references for metric threshold selection for metrics in Tables ~\ref{tab:acj_thresholds} and ~\ref{tab:aqj_thresholds}.}
\label{tab:metric_references}
\end{table}

\begin{table*}[ht]
\centering
\small
\begin{tabular}{|l|l|p{8cm}|}
\hline
\textbf{Category} & \textbf{Metric} & \textbf{Description} \\
\hline

Complexity & Cyclomatic Complexity (CC) & McCabe complexity per function; counts independent paths through code \\ \cline{2-3}
& Weighted Methods per Class (WMC) & Sum of CC for all methods in a class \\ \cline{2-3}
& Number of Methods (NOM) & Method count per class \\ \cline{2-3}
& Maintainability Index (MI) & Radon composite score combining CC, LOC, and Halstead volume \\
\hline

Size & LOC / SLOC / LLOC & Lines of code (total, source-only, logical) \\ \cline{2-3}
& Comments / Blank lines & Comment and blank line counts \\ \cline{2-3}
& Comment Ratio & Comments / SLOC \\ \cline{2-3}
& File count & Number of Python files \\ \cline{2-3}
& Class count & Number of classes \\ \cline{2-3}
& Function count & Standalone functions (not methods) \\ \cline{2-3}
& Method count & Methods inside classes \\ \cline{2-3}
& Component count & Packages/directories containing .py files \\
\hline

Unused Code & Unused Resource Count & Dead code: unused functions, classes, variables, imports (via vulture) \\ \cline{2-3}
& Unused Ratio & Unused elements / total defined elements \\
\hline

Cohesion & LCOM4 & Lack of Cohesion in Methods 4; connected components in method-attribute graph \\ \cline{2-3}
& Cohesive Ratio & Fraction of classes with LCOM4 = 1 \\
\hline

Dependency Graph & Coupling Density (CD) & $E / (N(N-1))$; ratio of actual to possible edges \\ \cline{2-3}
& Afferent Coupling (Ca) & Incoming dependencies per module (in-degree) \\ \cline{2-3}
& Efferent Coupling (Ce) & Outgoing dependencies per module (out-degree) \\ \cline{2-3}
& Graph Density & NetworkX density of the module dependency graph \\ \cline{2-3}
& Betweenness Centrality & Bridge/bottleneck identification per module \\ \cline{2-3}
& Avg Clustering Coefficient & Modular grouping tendency (on undirected projection) \\ \cline{2-3}
& Avg Shortest Path & Mean communication distance between reachable module pairs \\ \cline{2-3}
& Avg Weighted Degree & Mean interaction intensity (edge-weight-aware degree) \\ \cline{2-3}
& Dependency Concentration (Gini) & Gini coefficient on in-degree distribution \\ \cline{2-3}
& Q Value (Modularity) & Newman-Girvan modularity via greedy community detection \\ \cline{2-3}
& Instability & $Ce / (Ca + Ce)$ per module \\ \cline{2-3}
& Distance from Main Sequence & $|Abstractness + Instability - 1|$ per module \\ \cline{2-3}
& Total Dependencies & Internal + external edge count and ratio \\ \cline{2-3}
& External Usage Ratio & External imports / total imports \\ \cline{2-3}
& Autonomy Ratio & Internal calls / (internal + external) calls per module \\
\hline

Object-Oriented & DAM (Data Access Metric) & Encapsulation: (private + protected attrs) / total attrs \\ \cline{2-3}
& ANA / DIT (Avg Number of Ancestors) & Average depth of inheritance tree across classes \\ \cline{2-3}
& NOH (Number of Hierarchies) & Count of root classes that have children \\ \cline{2-3}
& MOA (Measure of Aggregation) & Count of attributes with user-defined types (composition) \\ \cline{2-3}
& MFA (Measure of Functional Abstraction) & Inherited methods / total methods \\ \cline{2-3}
& NOP (Number of Polymorphic Methods) & Count of methods overriding a parent method \\ \cline{2-3}
& CIS (Class Interface Size) & Public non-dunder method count \\ \cline{2-3}
& Extensibility & Ratio of classes with extension points (abstract methods, hooks) \\
\hline

Composite / Evolution & Evolvability & Weighted composite of extensibility, maintainability, low coupling, cohesion \\ \cline{2-3}
& Maintainability (composite) & Weighted composite of low complexity, low coupling, cohesion \\ \cline{2-3}
& Change Scenario Robustness & $1 - (\text{avg transitive impact} / \text{total components})$ \\
\hline

Patch-Specific & Patch Size & Lines added, removed, net; files modified \\ \cline{2-3}
& Patch Spread & Unique files and packages touched \\ \cline{2-3}
& Impact Zone & Transitive dependents of modified modules (size + ratio) \\ \cline{2-3}
& Touched Function CC & Cyclomatic complexity of added/modified functions \\ \cline{2-3}
& Cohesion Integration & LCOM4 delta + whether new methods connect to existing class attrs \\ \cline{2-3}
& Import Changes & New internal/external imports added or removed \\ \cline{2-3}
& API Surface Change & Public methods added/removed \\ \cline{2-3}
& New Attribute Encapsulation (DAM) & Private+protected vs public ratio for newly added attributes \\
\hline

\end{tabular}
\caption{Description of the implemented software architecture quality metrics from ~\citep{silva_arch_quality_metrics} alongside patch-specific metrics.}
\label{tab:software_metric_definitions}
\end{table*}


 \begin{figure*}[t]
    \centering
    \includegraphics[width=\textwidth]{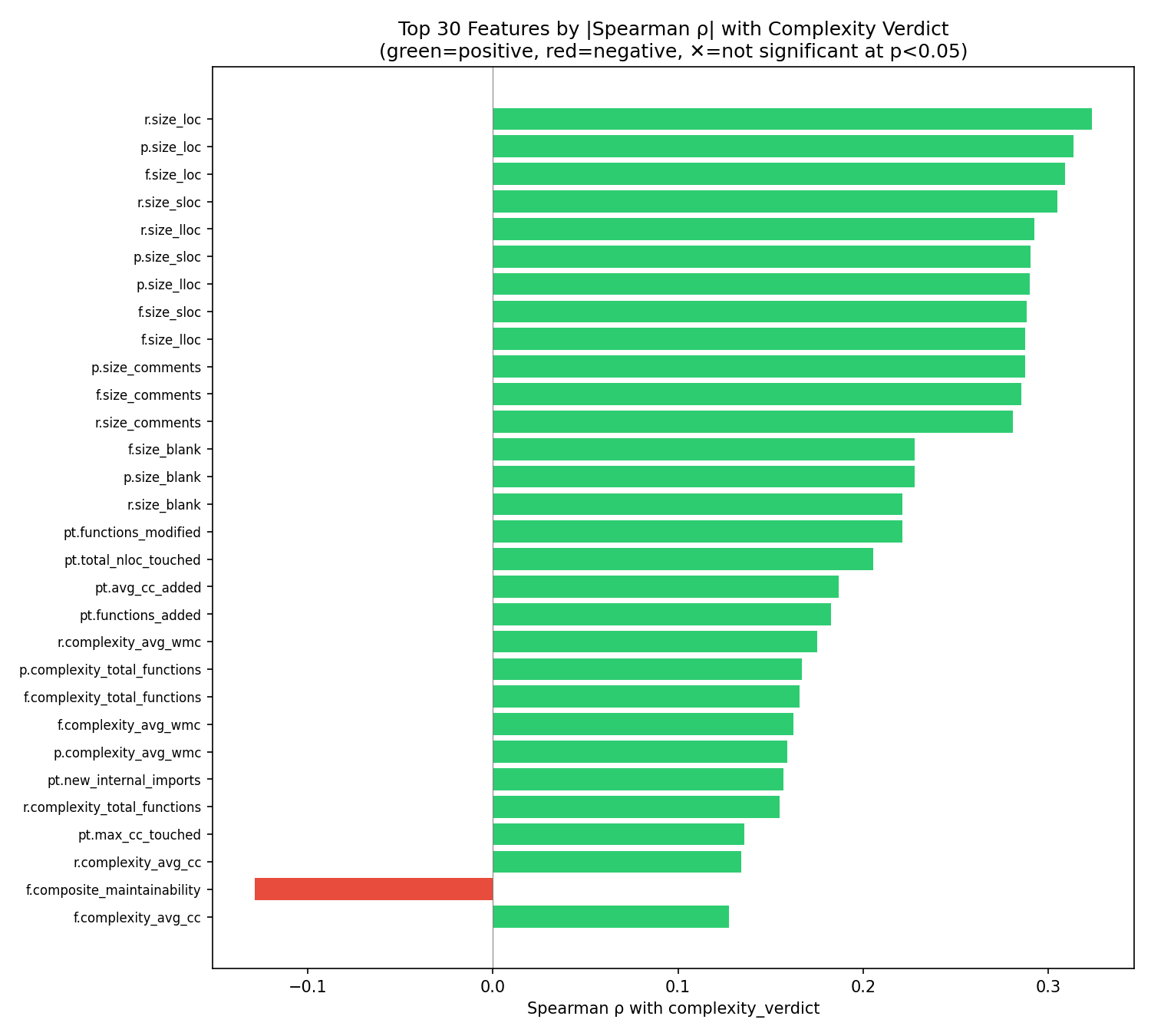}
    \caption{Spearman correlation of static metrics with ACJ scores (\textit{complexity\_verdict}), showing only ones with top-30 highest $|\rho|$; \textit{r}, \textit{p}, \textit{f}, and \textit{pt} metric prefix represent repo, package, file, and patch scope of the metric, respectively.}
    \label{fig:acj_score_correlation}
\end{figure*}

 \begin{figure*}[t]
    \centering
    \includegraphics[width=\textwidth]{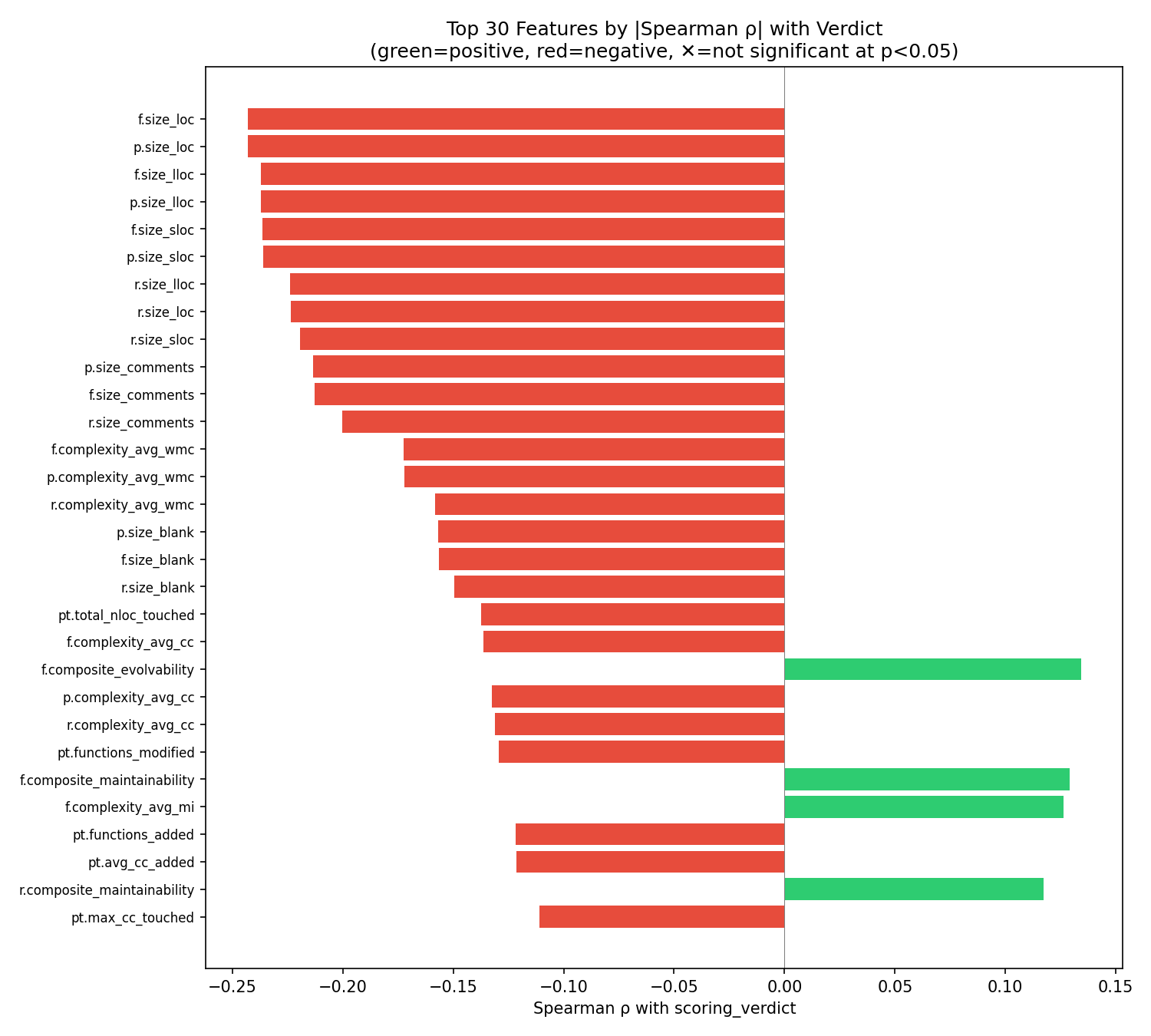}
    \caption{Spearman correlation of static metrics with AQJ scores (\textit{scoring\_verdict}), showing only ones with top-30 highest $|\rho|$; \textit{r}, \textit{p}, \textit{f}, and \textit{pt} metric prefix represent repo, package, file, and patch scope of the metric, respectively.}
    \label{fig:aqj_score_correlation}
\end{figure*}

\begin{table*}[h]
\centering
\caption{Five axes for architectural complexity assessment.}
\label{tab:complexity-axes}
\begin{tabular}{@{}p{2.8cm}p{5.5cm}p{4.5cm}@{}}
\toprule
\textbf{Axis} & \textbf{Definition} & \textbf{Scale} \\
\midrule
Scope of Context
& How many files, modules, or subsystems must be understood to arrive at the patch.
& minimal $\rightarrow$ single\_component $\rightarrow$ multiple\_components $\rightarrow$ system\_spanning \\[6pt]

Dependency Chain Depth
& How many architectural boundary crossings separate the change site from its necessary context.
& shallow (0--1 hops) $\rightarrow$ moderate (2--3) $\rightarrow$ deep (4+) \\[6pt]

Implicit Knowledge
& How much codebase-specific knowledge is required but not visible in the changed code itself.
& minimal $\rightarrow$ some\_conventions $\rightarrow$ architectural\_conventions $\rightarrow$ deep\_invariants \\[6pt]

Coordination Complexity
& Whether the edits and/or the reasoning require cross-boundary coordination, assessed separately at both levels.
& independent $\rightarrow$ locally\_coordinated $\rightarrow$ cross\_boundary $\rightarrow$ system\_wide \\[6pt]

Insight Density
& How hard it is to extract the relevant architectural understanding from the files, assuming the engineer already knows which files to read.
& low $\rightarrow$ moderate $\rightarrow$ high $\rightarrow$ expert \\
\bottomrule
\end{tabular}
\end{table*}

\end{document}